# Universal behavior of dense clusters of magnetic nanoparticles.

N A Usov[1,2] and O.N. Serebryakova[1,2]

[1]*National University of Science and Technology «MISIS», 119049, Moscow, Russia*
[2]*Pushkov Institute of Terrestrial Magnetism, Ionosphere and Radio Wave Propagation, Russian Academy of Sciences, (IZMIRAN) 142190, Troitsk, Moscow, Russia*

**Abstract**

A detailed numerical simulation of quasistatic hysteresis loops of dense clusters of interacting magnetic nanoparticles is carried out. Both clusters of magnetically soft and magnetically hard nanoparticles are considered. The clusters are characterized by an average particle diameter $D$, the cluster radius $R_c$, the particle saturation magnetization $M_s$, and the uniaxial anisotropy constant $K$. The number of particles in the cluster varies between $N_p$ = 30 - 120. The particle centers are randomly distributed within the cluster, their easy anisotropy axes being randomly oriented. It is shown that a rare assembly of identical random clusters of magnetic nanoparticles can be characterized by two dimensionless parameters: 1) the relative strength of magneto-dipole interaction, $K/M_s^2$, and the average particle concentration within the cluster, $\eta = VN_p/V_c$. Here $V$ is the nanoparticle volume, and $V_c$ is the volume of the cluster, respectively. In the strong interaction limit, $M_s\eta/H_a \gg 1$, where $H_a = 2K/M_s$ is the anisotropy field, the ultimate hysteresis loops of dilute assemblies of clusters have been constructed. In the variables ($M/M_s$, $H/M_s$) these hysteresis loops depend only on the particle volume fraction $\eta$. In the weak interaction limit, $M_s\eta/H_a \ll 1$, the assembly hysteresis loops in the variables ($M/M_s$, $H/H_a$) are close to the standard Stoner-Wohlfarth hysteresis loop.

## I. INTRODUCTION

Dense assemblies of magnetic nanoparticles show very rich and complex behavior because their properties depend on several important factors, such as the geometrical structure of the assembly, the nature of magnetic anisotropy of isolated nanoparticles, the presence of exchange and magneto- dipole interaction among the nanoparticles, etc. The long – range character of the inter-particle magneto-dipole interaction is the main problem that complicates significantly theoretical investigation of the dense nanoparticle assemblies. For comparison, the properties of assemblies of non interacting single-domain nanoparticles have been studied [1-10] in detail for nanoparticles with uniaxial, cubic and combined types of magnetic anisotropy. In addition, the behavior of dilute assemblies of superparamagnetic nanoparticles has been investigated depending on the ambient medium temperature and the value of the phenomenological damping constant [11 - 15]. However, theoretical study of dense assemblies of magnetic nanoparticles is more difficult as for an assembly of $N_p$ nanoparticles $N_p^2$ pair interactions must be taken into account. Therefore, the computational complexity increases rapidly with the number of the particles in the assembly.

Nevertheless, due to importance of the problem, the impact of a strong magneto-dipole interaction on the properties of dense nanoparticle assemblies have been investigated in many theoretical and experimental studies [16 - 41]. It seems reasonable to consider separately the magnetic nanoparticle assemblies of various geometrical structures, i.e. the ordered arrays of monodisperse magnetic nanoparticles [16-24], spatially disordered and randomly oriented assemblies of nanoparticles [25-34], and dense three-dimensional (3D) clusters of magnetic nanoparticles [35-41] that sometimes called multi-core nanoparticles. For ordered arrays of nanoparticles the existence of large magnetically ordered domains of nanoparticles separated by the domain walls has been revealed [19,20,23]. For spatially disordered assemblies the importance of the magneto- dipole interactions on basic magnetic characteristics of the assembly, such as the remanent magnetization [25,28], the hysteresis loop shape [27,32,33] and the relaxation properties [26, 28-34] have been demonstrated. The effective magnetic moments of dense clusters of magnetic nanoparticles in a low magnetic field and the magnetization curve of a dilute assembly of clusters in liquid were calculated [35-38] by means of Monte Carlo simulations. The magnetization relaxation process and magnetization reversal in 3D clusters of single domain nanoparticles were investigated [39-41] by solving Landau-Lifshitz-Gilbert (LLG) equation.

In spite of various interesting results obtained the understanding of the properties of dense assemblies of magnetic nanoparticles is still incomplete, since in general it is necessary to investigate the behavior of an assembly of interacting nanoparticles in solid matrix and viscous liquid, at moderate and high temperatures, under the influence of the quasi-static and alternating external magnetic field, etc. In this paper the quasi-static hysteresis loops of an assembly of magnetic nanoparticles distributed in a solid matrix are studied using numerical simulation under the following assumptions: 1) spherical nanoparticles of a radius $R$ are characterized by the saturation magnetization $M_s$ and uniaxial magnetic anisotropy constant $K$; 2) the temperature of the medium is much lower than the particle blocking temperature $T_b$, so that the thermal fluctuations of the particle magnetic moments can be neglected; 3) the particles of the assembly are contained in the quasi-spherical clusters of the given radius $R_{cl}$; 4) each cluster contains approximately the same number of nanoparticles $N_p \gg 1$. The assembly of clusters is sufficiently sparse, so that the magnetostatic interaction of nanoparticles of different clusters can be neglected. It is also assumed that the easy anisotropy axes of nanoparticles are oriented randomly, and the centers of nanoparticles occupy random positions within the cluster volume. However, the nanoparticles do not touch each other, so that the exchange interaction between neighboring nanoparticles is absent.

In spite of the constraints assumed, this model seems to be able to describe adequately the behavior the experimentally investigated assemblies of clusters with strong magneto- dipole interaction [35-40]. Note that even with the simplifications adopted, there are 5 independent parameters of the model, i.e. the saturation magnetization $M_s$, the uniaxial anisotropy constant $K$, the particle radius $R$, the cluster radius $R_{cl}$, and the total number of the nanoparticles in the cluster $N_p$. In this paper, based on the detailed numerical simulations, a unified description of the quasi-static hysteresis loops of the assembly of dense clusters of nanoparticles is obtained. The hysteresis loops of the assembly are averaged over a sufficient number (~ 20 - 30) of random cluster realizations. The total number of particles in the whole assembly is about of several thousands. This is sufficient to obtain statistically convincing results.

## II. Basic equations

Dynamics of the unit magnetization vector $\vec{\alpha}_i$ of the $i$-th nanoparticle of the cluster is described by the LLG equation

$$\frac{\partial \vec{\alpha}_i}{\partial t} = -\gamma [\vec{\alpha}_i, \vec{H}_{ef,i}] + \kappa \left[\vec{\alpha}_i, \frac{\partial \vec{\alpha}_i}{\partial t}\right], \quad i = 1, 2, N_p, \quad (1)$$

where $\kappa$ is the phenomenological damping constant and $\gamma$ is the gyromagnetic ratio. The effective magnetic field acting on a separate nanoparticle can be calculated as a derivative of the total cluster energy

$$\vec{H}_{ef,i} = -\frac{\partial W}{V M_s \partial \vec{\alpha}_i}. \quad (2)$$

Here $V = \pi D^3/6$ is the nanoparticle volume, $D = 2R$ is the particle diameter. The total magnetic energy of the cluster $W = W_a + W_Z + W_m$ is a sum of the magneto-crystalline anisotropy energy $W_a$, Zeeman energy $W_Z$ of the particles in the external uniform magnetic field $\vec{H}$, and the energy of mutual magneto-dipole interaction of the particles $W_m$.

For nanoparticles of nearly spherical shape with uniaxial type of magneto-crystalline anisotropy the magneto-crystalline anisotropy energy is given by

$$W_a = KV \sum_{i=1}^{N_p} \left(1 - (\vec{\alpha}_i \vec{e}_i)^2\right), \quad (3)$$

where $\vec{e}_i$ is the orientation of the easy anisotropy axis of $i$-th particle of the cluster. Zeeman energy $W_Z$ of the cluster in the external uniform magnetic field is given by

$$W_Z = -M_s V \sum_{i=1}^{N_p} (\vec{\alpha}_i \vec{H}) \quad (4)$$

Next, for spherical uniformly magnetized nanoparticles the magnetostatic energy of the cluster can be represented as the energy of the point interacting dipoles. Let $\vec{r}_i$ be the coordinates of the particle centers within the cluster. Then the magneto-dipole interacting energy reads

$$W_m = \frac{M_s^2 V^2}{2} \sum_{i \neq j} \frac{\vec{\alpha}_i \vec{\alpha}_j - 3(\vec{\alpha}_i \vec{n}_{ij})(\vec{\alpha}_j \vec{n}_{ij})}{|\vec{r}_i - \vec{r}_j|^3}, \quad (5)$$

where $\vec{n}_{ij}$ is the unit vector along the line connecting the centers of $i$-th and $j$-th particles, respectively.

Thus, the effective magnetic field acting on the $i$-th nanoparticle of the cluster is given by

$$\vec{H}_{ef,i} = H_a (\vec{\alpha}_i \vec{e}_i) \vec{e}_i + \vec{H} + M_s V \sum_{j \neq i} \frac{\vec{\alpha}_j - 3(\vec{\alpha}_j \vec{n}_{ij}) \vec{n}_{ij}}{|\vec{r}_i - \vec{r}_j|^3} \quad (6)$$

where $H_a = 2K/M_s$ is the particle anisotropy field.

We start the calculation of a quasistatic hysteresis loop of a cluster at a sufficiently high value of the external uniform magnetic field sufficient for assembly saturation. Therefore, in the initial state the unit magnetization vectors of all particles are set in the direction parallel to the applied magnetic field. At the next step the magnetic field decreases by a small value, $dH = 1 - 2$ Oe, and evolution of the initial magnetization state is calculated according to Eqs. (1) – (5). For a cluster consisting of several tens of nanoparticles it is necessary to make usually $10^4 - 10^5$ iterations with a sufficiently small time increment to find stable magnetization state of a cluster at a given value of the external magnetic field. In accordance with the Eq. (1), the final magnetization state is assumed to be stable under the condition

$$\max_{(1 \leq i \leq N_p)} |[\vec{\alpha}_i, \vec{H}_{ef,i}]| < 10^{-8}, \quad (7)$$

which means that the unit magnetization vector of each nanoparticle is parallel with a sufficiently high accuracy to the effective magnetic field acting on this nanoparticle.

In this manner, for a given set of initial material and geometrical parameters, i.e. $K$, $M_s$, $R$, $R_{cl}$ and $N_p$, one can obtain a quasi-static hysteresis loop of a nearly spherical cluster of nanoparticles, which depend, generally speaking, on the set of the coordinates of the nanoparticle centers $\{r_i\}$, as well as on the set the orientations $\{e_i\}$ of the particle easy anisotropy axes. However, the calculations show that in the limit $N_p \gg 1$ the hysteresis loops obtained for different realizations of random variables $\{r_i\}$ and $\{e_i\}$ differ only slightly from each other.

Moreover, if one assume an assembly of non-interacting identical clusters, and calculate a common hysteresis loop of the assembly averaged over a sufficiently large number of random cluster realizations, one obtains a non-random hysteresis loop. The latter characterizes the behaviour of the assembly of non-interacting random nanoparticle clusters. Calculations show that in the limit $N_p \gg 1$ the averaged hysteresis loop of cluster assembly has a rather small dispersion even being averaged over 20 - 30 independent realizations of random clusters with the fixed values of the initial parameters $K$, $M_s$, $R$, $R_{cl}$ and $N_p$.

The question naturally arises, how to describe these non-random hysteresis loops and to establish an actual set of dimensionless parameters which determine the averaged hysteresis loop behaviour. Note that the average value of the effective magnetic field, equation (6), acting on a typical particle of the cluster, can be qualitatively represented in the form

$$\left\langle \vec{H}_{ef} \right\rangle = H_a \left\langle \frac{\partial w_a}{\partial \vec{\alpha}} \right\rangle + \vec{H}_0 + M_s \eta \left\langle \frac{\partial w_m}{\partial \vec{\alpha}} \right\rangle. \quad (8)$$

Here $\eta = V N_p / V_{cl}$ is a geometric parameter characterizing the packing density of magnetic nanoparticles within the cluster volume, $V_{cl} = 4\pi R_{cl}^3 / 3$. The dimensionless values

$$\left\langle \frac{\partial w_a}{\partial \vec{\alpha}} \right\rangle = \left\langle (\vec{\alpha}_i \vec{e}_i) \vec{e}_i \right\rangle$$

$$\left\langle \frac{\partial w_m}{\partial \vec{\alpha}} \right\rangle = \left\langle \frac{4\pi}{3 N_p} \sum_{j \neq i} \frac{\vec{\alpha}_j - 3(\vec{\alpha}_j \vec{n}_{ij}) \vec{n}_{ij}}{\left| (\vec{r}_i - \vec{r}_j) / R_{cl} \right|^3} \right\rangle$$

characterize the intensity of the anisotropy and demagnetization fields acting on a typical particle of the cluster. Let us normalize the averaged effective magnetic field (8) by the value of the particle anisotropy field

$$\left\langle \frac{\vec{H}_{ef}}{H_a} \right\rangle = \left\langle \frac{\partial w_a}{\partial \vec{\alpha}} \right\rangle + \frac{\vec{H}_0}{H_a} + \frac{M_s \eta}{H_a} \left\langle \frac{\partial w_m}{\partial \vec{\alpha}} \right\rangle. \quad (9)$$

From this equation it is clear that the influence of the magneto- dipole interaction on the properties of random clusters of magnetic nanoparticles depends on the product of the dimensionless ratio $M_s/H_a$ and the filling factor $\eta < 1$. It is well known [3], that in the limit of weak magneto- dipole interaction, $M_s \eta / H_a \ll 1$, all random assemblies are described by the universal Stoner - Wohlfarth hysteresis loop. On the other hand, in the limit of strong magneto- dipole interaction, $M_s \eta / H_a \gg 1$, it is reasonable to normalize the effective magnetic field (8) by the particle saturation magnetization $M_s$. Then one obtains the relation

$$\left\langle \frac{\vec{H}_{ef}}{M_s} \right\rangle = \frac{H_a}{M_s} \left\langle \frac{\partial w_a}{\partial \vec{\alpha}} \right\rangle + \frac{\vec{H}_0}{M_s} + \eta \left\langle \frac{\partial w_m}{\partial \vec{\alpha}} \right\rangle \approx \frac{\vec{H}_0}{M_s} + \eta \left\langle \frac{\partial w_m}{\partial \vec{\alpha}} \right\rangle \quad (10)$$

Based on the approximate equation (10) one can hypothesize that in the limit of strong magneto- dipole interaction, $M_s \eta / H_a \gg 1$, there are ultimate hysteresis loops of the cluster assembly that in variables ($M/M_s$, $H_0/M_s$) depend only on the value of the cluster filling factor $\eta$.

If this hypothesis is correct, the properties of an assembly of non-interacting clusters with a sufficiently large number of particles, $N_p \gg 1$, depend mainly on two dimensionless parameters, i.e. the ratio $K/M_s^2$, which is determined only by the material parameters of nanoparticles, and the filling factor $\eta$ characterizing the nanoparticle density in the volume of the quasi-spherical cluster.

### III. Results and discussion

To check the hypothesis stated above it is necessary to analyze the shape of the averaged hysteresis loops of a dilute assembly of nanoparticle clusters depending on the set of dimensionless parameters ($\eta$, $K/M_s^2$). To do so, we carried out a detailed calculation of the averaged hysteresis loops of the clusters with different sets of initial parameters $K$, $M_s$, $D$, $R_{cl}$ and $N_p$. In the simulations performed the material parameters of the nanoparticles vary in the intervals $K = 5\times10^4 – 10^5$ erg/cm$^3$, and $M_s = 300 – 800$ emu/cm$^3$, respectively. Thus, the dimensionless parameter $K/M_s^2$ varies within the range $0.08 \leq K/M_s^2 \leq 1.1$. The diameters of the nanoparticles studied are in the range of $D = 20 - 40$ nm, the number of particles in the cluster is given by $N_p = 30 - 120$, the cluster density being $\eta < 0.5$.

Random clusters with a given number of particles $N_p$ were created in this study as follows. First, we generated dense enough and approximately uniform set of $N$ random points $\{\rho_i\}$ within a spherical volume with the radius $R_{cl}$ so that $|\rho_i| \leq R_{cl}$, $i = 1,2, ... N$, $N \gg N_p$. Center of the first nanoparticle was placed in the first random point, $r_1 = \rho_1$. Then all random points with coordinates $|\rho_i - r_1| \leq 2R$ were removed from the initial set of the random points. After this operation, any point in the remaining set of random points could serve as a center of the second nanoparticle. For example, one can put simply $r_2 = \rho_2$. At the next step one removes all random points whose coordinates satisfy the inequality $|\rho_i - r_2| \leq 2R$. This procedure is repeated until all $N_p$ nanoparticle centers are placed within the cluster volume. As a result, all random nanoparticle centers lie within a sphere of radius $R_{cl}$, so that $|r_i| \leq R_{cl}$, $i = 1, 2, .. N_p$. Furthermore, none of the nanoparticles is in a direct contact with the neighboring nanoparticles.

This algorithm enables one to construct random quasi- spherical clusters of magnetic nanoparticles for moderate values of the cluster filling factor $\eta < 0.5$. The orientations of the easy anisotropy axes of uniaxial nanoparticles $\{e_i\}$ were also chosen randomly and independently in the upper hemisphere of the spherical coordinates, $0 < \theta < \pi/2$.

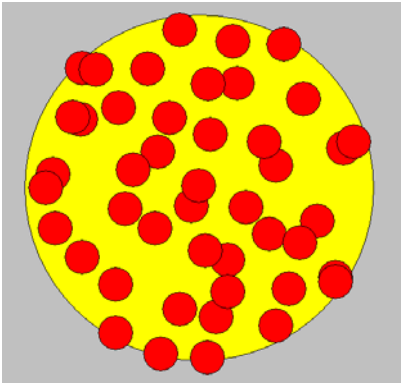

Figure 1. An example of a random quasi-spherical cluster of nanoparticles of radius $R_{cl}$ = 320 nm, comprising $N_p$ = 120 nanoparticles of radius $R$ = 20 nm. The filling cluster density is given by $\eta = VN_p/V_{cl} = 0.23$.

Figure 1 shows an example of quasi- spherical random cluster of nanoparticles created using the algorithm described above. Only particles whose centers are located in a layer $|z| < R$ near the equatorial plane $z = 0$ are shown. The overlap of some particles, visible in figure 1, is only apparent. In fact the particles are at different distances from the plane $z = 0$, and do not intersect with each other.

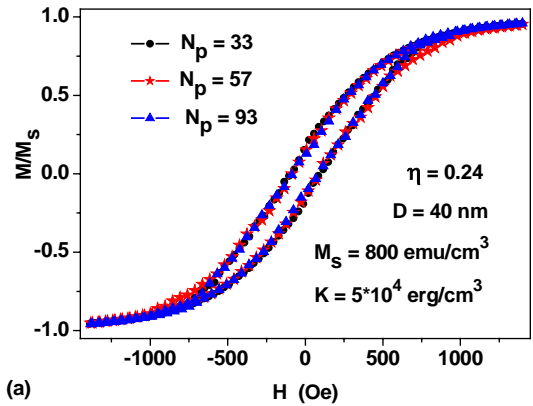

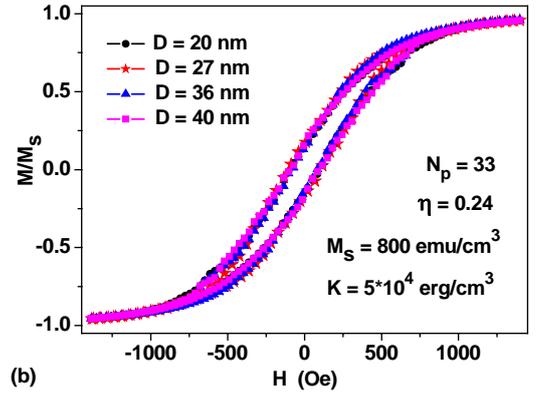

Figure 2. Averaged hysteresis loops of an assembly of noninteracting clusters with a given particle packing density $\eta$ = 0.24: a) clusters with different numbers of particles of a fixed diameter $D$ = 40 nm; b) clusters of particles of different diameters, $D$ = 20 - 40 nm, but with a fixed number of particles $N_p$ = 33.

Figure 2 shows the averaged hysteresis loops of dilute assemblies of clusters with the same filling factor $\eta$, but having different geometric structures. For each set of parameters the loop averaging is done over 20 independent realizations of random clusters. Figure 2a shows the hysteresis loop of clusters with different amounts of particles of the same diameter. The volume of the spherical cluster varies inversely to the particle number, $V_{cl} \sim 1/N_p$, to keep the filling factor $\eta$ unchanged. In figure 2b, conversely, the particle diameters are different, but number of particles in the cluster is maintained constant. The volumes of the clusters are also changed so to keep the $\eta$ value unchanged. As figures 2a and 2b show, in both cases the shape of the averaged hysteresis loops of the assemblies studied is almost the same. Such insensitivity of the averaged hysteresis loops on the number of particles in the cluster in the limit $N_p \gg 1$, or on the particle diameters, have been observed for other filling factors investigated.

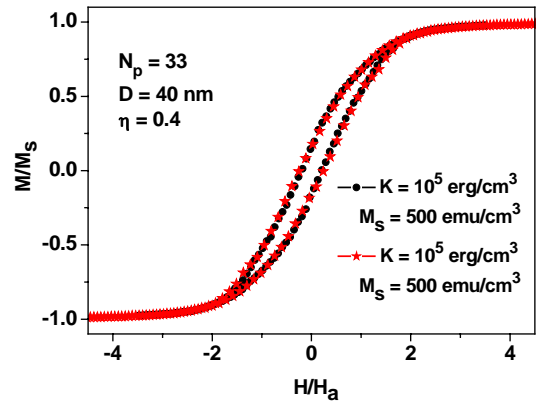

Figure 3. A comparison of the average hysteresis loops for rare assembles of clusters with different values of $K$ and $M_s$, but with the same ratio $K/M_s^2 \approx 0.4$.

In figure 3 we compare the averaged hysteresis loops of rare assemblies of clusters with the same filling factor $\eta$ = 0.4 and the same nanoparticle diameter $D$ = 40 nm, but with different values of the material constants $K$ and $M_s$. For the first assembly with $K = 10^5$ erg/cm$^3$ and $M_s$ = 500 emu/cm$^3$ the dimensionless ratio is given by $K/M_s^2$ = 0.4. For the second assembly with $K = 5\times10^4$ erg/cm$^3$ and $M_s$ = 350 emu/cm$^3$ one has nearly the same value, $K/M_s^2$ = 0.41. The hysteresis loops in figure 3 are normalized to the corresponding anisotropy fields, $H_{a1}$ = 400 Oe and $H_{a2}$ = 285.7 Oe, for the first and second assemblies, respectively. As figure 3 shows, due to this normalization the calculated hysteresis loops coincide with sufficient accuracy. Note that under the condition $K/M_s^2$ = const the anisotropy field ratio equals to the ratio of the saturation magnetizations, $H_{a1}/H_{a2} = M_{s1}/M_{s2}$. Therefore, the loops in figure 3 will also coincide being normalized on the saturation magnetizations $M_{s1}$ and $M_{s2}$ of the nanoparticles, respectively.

Next, we study the dependence of the averaged hysteresis loops on the parameter $K/M_s^2$ at a fixed cluster packing density $\eta$. As figure 4a shows, at a fixed packing density $\eta$ and with increasing $K/M_s^2$ ratio the averaged hysteresis loops of dilute assembly of clusters in variables ($M/M_s$, $H/H_a$) is gradually approaching the standard Stoner- Wohlfarth [3] hysteresis loop for non-oriented assembly of spherical single-domain nanoparticles with uniaxial anisotropy.

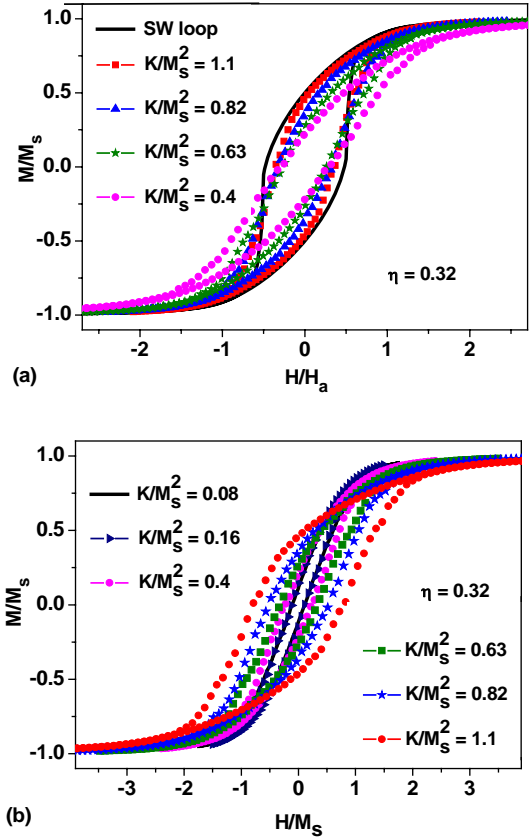

Figure 4. The evolution of the averaged hysteresis loops as a function of the ratio $K/M_s^2$ for fixed filling factor $\eta$ = 0.32: a) approaching the standard Stoner- Wohlfarth hysteresis loop with increasing $K/M_s^2$ ratio; b) approaching the ultimate hysteresis loop with decreasing $K/M_s^2$ ratio.

On the other hand, with decreasing of $K/M_s^2$ ratio the averaged hysteresis loop of a dilute assembly of clusters evolves in the opposite direction. As shown in figure 4b, with a reduction of the $K/M_s^2$ ratio the hysteresis loop area decreases and in the variables ($M/M_s$, $H/M_s$) the loops gradually stretched approaching to an ultimate hysteresis loop, characteristic for clusters with strong magneto- dipole interaction. Indeed, as shown in figure 4b, the shape of the hysteresis loop is almost unchanged for values $K/M_s^2$ < 0.2. Similar evolution of the hysteresis loops depending on the ratio $K/M_s^2$ has been observed for other values of filling factor $\eta$ = 0.06, 0.13, 0.24, 0.4 and 0.49 studied. As we shall see, in accordance with the hypothesis stated in Section II, the slope and area of the ultimate hysteresis loops are determined mainly by the cluster volume fraction $\eta$.

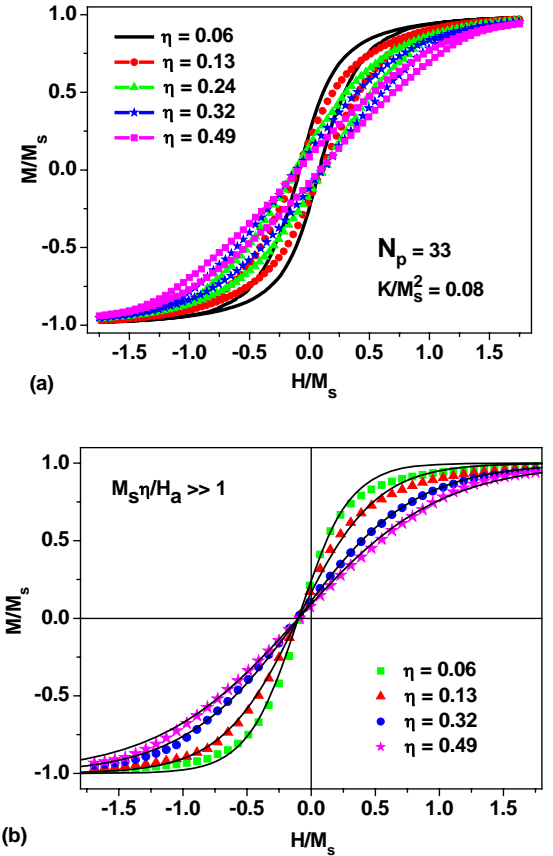

Figure 5. a) The dependence of the average ultimate hysteresis loops on the filling factor $\eta$. b) analytical approximation of the ultimate hysteresis loops: dots are the result of numerical simulation, solid lines are drawn according to equation (11).

Figure 5a shows the averaged hysteresis loops of dilute assembles of clusters of nanoparticles in the limit of strong magneto- dipole interaction, $M_s\eta/H_a \gg 1$, as a function of the cluster filling factor $\eta$. As shown in Figure 5b, the descending branches of these ultimate hysteresis loops can be described with a satisfactory accuracy by the simple formula

$$\langle M \rangle = M_s \tanh\left(\frac{H + H_c}{M_s f(\eta)}\right). \qquad (11)$$

The coercive force, $H_c = 0.1M_s$, turns out to be independent of the parameter $\eta$, the function $f(\eta)$ being $f(\eta) = 0.26+2.7\eta -2.1\eta^2$.

It is also important to determine in which area of the dimensionless parameters $\eta$ and $K/M_s^2$ the hysteresis loops are close to standard Stoner- Wohlfarth hysteresis loop, and for what values of these parameters the loops are close to the ultimate hysteresis loops shown in Fig. 5. To this end, for some values of $\eta$ = 0.06, 0.13, 0.24, 0.32, 0.4 and 0.49, we calculated the averaged hysteresis loops of dilute assemblies of clusters of nanoparticles with different ratios $K/M_s^2$ within the range $0.08 \leq K/M_s^2 \leq 0.82$.

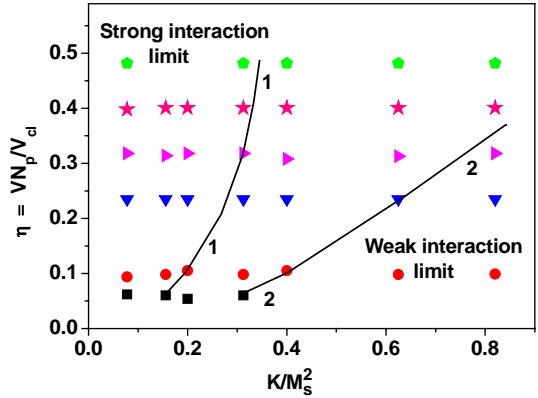

Figure 6. Areas of weak and strong magneto-dipole interaction of monodispersed magnetic nanoparticles on the parameter plane ($\eta$, $K/M_s^2$).

The numerical simulation of the averaged hysteresis loops of dilute assemblies of clusters of magnetic nanoparticles has been performed for the different pairs of dimensionless parameters ($\eta$, $K/M_s^2$) shown in figure 6. As numerical simulations show, to the left of the curve 1-1 in figure 6 the averaged hysteresis loops are close to the ultimate loops shown in figure 5. The latter are realized in the strong magneto-dipole interaction, $M_s\eta/H_a \gg 1$. On the other hand, to the right of the curve 2-2 in figure 6 the hysteresis loops are close to the standard Stoner – Wohlfarth hysteresis loop [3] for non-oriented assembly of non-interacting single-domain nanoparticles with uniaxial magnetic anisotropy.

### IV. Conclusions

In this paper, using numerical simulation the averaged hysteresis loops of dilute assemblies of random clusters of uniaxial single-domain magnetic nanoparticles have been investigated. The assemblies studied are characterized by a set of material, $K$ and $M_s$, and geometric, $D$, $R_{cl}$, $N_p$ parameters. It is shown that the properties of dilute assembly of identical random clusters of nanoparticles are determined by two dimensionless parameters, i.e. the ratio $K/M_s^2$, and the average particle concentration within the cluster, $\eta = VN_p/V_{cl}$. In the strong interaction limit, $M_s\eta/H_a \gg 1$, the ultimate hysteresis loops of dilute assemblies have been constructed. In the variables ($M/M_s$, $H/M_s$) these hysteresis loops depend only on the particle concentration, which varies in the range $\eta < 0.5$. On the other hand, in the weak interaction limit, $M_s\eta/H_a \ll 1$ the assembly hysteresis loops are close to the standard Stoner-Wohlfarth hysteresis loop of non-oriented assembly of uniaxial single-domain particles.

Note that some of the restrictions of the model considered in this paper can be safely avoided. Indeed, initially it was assumed that each isolated cluster contains approximately the same number of magnetic nanoparticles $N_p$. But then it was found that the properties of the individual clusters depend actually on the cluster volume fraction $\eta$. Therefore, the results obtained are also valid for the case of non-interacting clusters with different numbers of the nanoparticles provided by the clusters have nearly the same volume fraction $\eta$. Further, the model assumes that the clusters of the nanoparticles have a quasi-spherical shape. Apparently, the properties of clusters of ellipsoidal shape can be described in a similar way, taking into account the change in the average demagnetizing field existing within the cluster. The model considered in this paper allows generalizations in various directions. It would be desirable to study the changes that occur when one takes into account a distribution of the nanoparticles diameters. It would be interesting also to consider the case when a certain amount of nanoparticles of the cluster can be in direct contact, so that the exchange coupling exists among these nanoparticles. However, these issues deserve special consideration.

### Acknowledgments

The authors wish to acknowledge the financial support of the Ministry of Education and Science of the Russian Federation in the framework of Increase Competitiveness Program of NUST «MISIS», contract № K2-2015-018.